\documentclass{article}

\usepackage{PRIMEarxiv}

\usepackage[utf8]{inputenc} 
\usepackage[T1]{fontenc}    
\usepackage{hyperref}       
\usepackage{url}            
\usepackage{booktabs}       
\usepackage{amsfonts}       
\usepackage{nicefrac}       
\usepackage{microtype}      
\usepackage{lipsum}
\usepackage{amsmath}
\usepackage{fancyhdr}       
\usepackage{graphicx}       
\usepackage{algorithm}
\usepackage{algpseudocode}
\usepackage{multirow}
\usepackage{authblk}        

\graphicspath{{media/}}     

\title{In-context learning for medical image segmentation
}

\author[1,2,3]{Eichi Takaya}
\author[1]{Shinnosuke Yamamoto}
\affil[1]{Department of Diagnostic Imaging, Tohoku University Graduate School of Medicine, Miyagi, Japan \thanks{\texttt{eichi.takaya.d5@tohoku.ac.jp}}}
\affil[2]{AI Lab, Tohoku University Hospital, Miyagi, Japan}
\affil[3]{ Department of Medical Information and Communication Technology Research, St. Marianna University Graduate School of Medicine, Kanagawa, Japan}

\begin{document}
\maketitle

\begin{abstract}
Accurate annotation of medical images, such as MRI and CT scans, is critical for assessing treatment efficacy and planning radiotherapy. However, the substantial time and effort required of medical professionals limit the volume of annotated data, thereby constraining AI applications in medical imaging. To address this challenge, we propose in-context cascade segmentation (ICS), a novel method that minimizes annotation needs while maintaining high segmentation accuracy for sequential medical images. ICS extends the UniverSeg framework, which performs few-shot segmentation based on support images without additional training. By iteratively incorporating the inference results of each slice into the support set, ICS propagates information forward and backward through the sequence, ensuring inter-slice consistency. We evaluate ICS on the HVSMR dataset encompassing eight cardiac regions. Experimental results demonstrate that ICS significantly outperforms baseline methods, particularly in complex anatomical regions where maintaining boundary consistency across slices is crucial. Furthermore, we show that the number and placement of initial support slices can markedly influence segmentation accuracy. ICS thus provides a promising solution for reducing annotation burdens while delivering robust segmentation results, paving the way for broader implementation in both clinical and research settings.
\end{abstract}


\section{Introduction}
Research on AI applications in healthcare, particularly automated segmentation in medical imaging, has accelerated in recent years in tandem with deep learning advances \cite{mall_comprehensive_2023}\cite{azad_medical_2024}. Segmentation of sequential imaging data, such as CT and MRI scans, is essential for surgical and radiotherapy planning \cite{dot_fully_2022}\cite{harrison_machine_2022}, as well as for evaluating cancer treatment efficacy \cite{zhang_deeprecs_2022}. Given the high manual workload and associated costs, there is a pressing need to automate these processes.

In addition, building AI-assisted diagnostic systems frequently requires defining regions of interest (ROIs) for lesions within training datasets. For instance, ROIs are used for image cropping in classification tasks \cite{shimokawa_deep_2023} and for producing radiomics features \cite{haraguchi_radiomics_2023}. Furthermore, recent findings suggest that performing pixel-level lesion classification, differentiating benign from malignant conditions using segmentation methods, can improve diagnostic accuracy \cite{hooper_case_nodate}.

A range of deep neural network architectures has been introduced, starting with U-Net \cite{ronneberger_u-net_2015}, More recently, vision-transformer-based models have gained popularity \cite{cao_swin-unet_2022}. Large-scale pretraining on datasets such as ImageNet \cite{das_attention-unet_2024} and RadImageNet \cite{mei_radimagenet_2022}\cite{xu_vis-mae_2025}, alongside foundation models capable of zero-shot inference \cite{wu_medical_2023}\cite{zhu_medical_2024}, has further enhanced generalization performance across diverse segmentation tasks. Nonetherless, domain shifts (differences in patient populations or imaging protocols), modality shifts (variations between imaging modalities such as CT and MRI), and site shifts (differences in imaging equipment or procedures across institutions) remain challenging \cite{ouyang_causality-inspired_2023}\cite{guan_domain_2022}\cite{zhang_generalizing_2020}. Additionally, these approaches often exhibit limited generalizability to rare diseases \cite{hasani_artificial_2022}, frequently necessitating the creation of new, task-specific training sets. Reducing the burden of annotating entire image sequences therefore remains a critical objective.

Our study aims to develop a method that minimizes the annotation load for sequential imaging datasets. We introduce a segmentation approach based on in-context learning, specifically leveraging UniverSeg \cite{butoi_universeg_2023}, a framework designed for few-shot medical image segmentation. By iteratively incorporating minimal initial labels (“support” images) and augmenting the support set with predicted masks, we implement an in-context cascade segmentation (ICS) strategy. We evaluated ICS on eight cardiac segmentation tasks in the HVSMR dataset \cite{pace_hvsmr-20_2024}, highlighting its performance and characteristics.

The remainder of this paper is organized as follows. Section 2 reviews related work on medical image segmentation, including in-context and semi-supervised learning approaches. Section 3 describes our problem setting and details the proposed method. Section 4 outlines the experimental setup and reports results on HVSMR data. Section 5 discusses the findings, and Section 6 concludes with recommendations for future research directions and open challenges.

\section{Related Works}
\subsection{Medical Image Segmentation}
Prior to the widespread adoption of deep learning, segmentation in medical imaging was predominantly approached through traditional machine learning methods employing both unsupervised and supervised strategies \cite{pham_current_2000}. In unsupervised methods, clustering and region-based segmentation algorithms, such as K-means and watershed, were extensively utilized \cite{ng_medical_2006}\cite{ng_masseter_2008}. These approaches partitioned images based on the statistical properties of pixels or voxels, eliminating the need for prior labeling. Supervised techniques typically relied on engineered features and models such as support vector machines (SVMs) and random forests for pixel-level classification \cite{ricci_retinal_2007}\cite{mahapatra_analyzing_2014}. Although these methods succeeded in certain tasks, they generally lacked accuracy and scalability.

Following the work by Ciresan et al. \cite{ciresan_deep_2012}, deep neural networks enabled automatic feature extraction from images, thereby facilitating segmentation. Subsequently, U-Net \cite{ronneberger_u-net_2015} emerged as a standard architecture, leveraging an encoder–decoder design and skip connections to capture both local and global features for high-precision segmentation. In recent years, transformer-based models and 3D segmentation techniques have further advanced the state of the art by tackling complex anatomical structures and diverse imaging modalities \cite{cao_swin-unet_2022}\cite{chen_transunet_2024}\cite{milletari_v-net_2016}. Large-scale segmentation frameworks, such as MedSAM \cite{wu_medical_2023}\cite{zhu_medical_2024} built on Segment anything model (SAM)\cite{kirillov_segment_2023}\cite{ravi_sam_nodate}, have also shown promise in achieving high accuracy from minimal annotations. The UniverSeg \cite{butoi_universeg_2023} model used in our study belongs to this class of large-scale medical image segmentation frameworks.

\subsection{In-context Learning}
In-context learning \cite{dong_survey_2024}\cite{min_rethinking_2022} was initially introduced in large language models (LLMs), wherein a pre-trained model can adapt to new tasks given only a few examples without requiring additional training. In GPT-3 \cite{brown_language_2020}, which first popularized the term, it was demonstrated that just a small number of examples in the prompt can enable the model to generalize effectively. Moreover, theoretical work by Oswald et al. \cite{oswald_transformers_2023} suggests that the underlying mechanism in transformers performing in-context learning parallels gradient descent.

The concept of in-context learning has been extended to image recognition \cite{zhang_what_2023}. In UniverSeg \cite{butoi_universeg_2023}, for example, labeled "support" images guide the segmentation of new query images. In this study, we further develop this approach of in-context learning and apply it as ICS, where inference results are sequentially utilized to improve segmentation consistency across sequential slices.

\subsection{Semi-supervised Learning}
Semi-supervised learning seeks to leverage unlabeled data when labeled datasets are scarce, making it especially relevant in medical image segmentation, where annotation is both expensive and time-consuming \cite{van_engelen_survey_2020}\cite{jiao_learning_2024}. Common approaches include self-training \cite{chaitanya_local_2023}, co-training \cite{xia_uncertainty-aware_2020}, and graph-based methods \cite{wang_graphcl_2024}. Self-training re-trains the model using pseudo-labels generated from unlabeled data, while co-training employs two distinct classifiers that iteratively learn from each other’s pseudo-labels. In graph-based methods, data are represented as a graph, and labels are propagated to unlabeled nodes.

Among these techniques, sequential semi-supervised segmentation (4S) \cite{takaya_sequential_2021} is particularly related to our approach. It begins with a small number of consecutively labeled slices and gradually extends the labeled region to neighboring slices by treating inferred masks as pseudo-labels. However, 4S involves iterative re-training, incurring considerable computational costs. By contrast, our method capitalizes on in-context learning to update the support set with inference results, avoiding explicit parameter re-training. This strategy reduces both computational overhead and development time while maintaining robust segmentation performance.

\section{Methods}
\subsection{Problem Definition}

In this study, we consider a volumetric dataset
\[
V = \{\text{slice}_1, \text{slice}_2, \ldots, \text{slice}_n\},
\]
where each \(\text{slice}_k\) (\(1 \le k \le n\)) is a two-dimensional cross-sectional image (e.g., from CT or MRI). Among these \(n\) slices, only \(m\) slices have corresponding segmentation masks (i.e., ground-truth labels). We define this collection of labeled slices as the support set \(S\):
\[
S 
= \Bigl\{
    \bigl(\text{slice}_{\ell_1},\, \text{label}_{\ell_1}\bigr),
    \bigl(\text{slice}_{\ell_2},\, \text{label}_{\ell_2}\bigr),
    \ldots,
    \bigl(\text{slice}_{\ell_m},\, \text{label}_{\ell_m}\bigr)
\Bigr\},
\]
where \(\ell_1, \ell_2, \ldots, \ell_m\) are the indices of the labeled slices, satisfying
\[
1 \,\le \ell_1 < \ell_2 < \cdots < \ell_m \,\le n.
\]
Each \(\text{label}_{\ell_k}\) is a segmentation mask that indicates the ground-truth regions for \(\text{slice}_{\ell_k}\).

The objective of this work is to generate segmentation masks for the remaining \((n-m)\) unlabeled slices, namely those with indices
\[
k \in \{1, 2, \ldots, n\} \,\setminus\, \{\ell_1, \ell_2, \ldots, \ell_m\}.
\]
For each unlabeled slice \(\text{slice}_k\), we denote the predicted segmentation mask by \(\hat{M}_k\), where
\[
\hat{M}_k = f\bigl(\text{slice}_k,\; S\bigr).
\]
Here, \(f\) is a pretrained model (e.g., UniverSeg) that takes both the query slice \(\text{slice}_k\) and the support set \(S\) as input. The output \(\hat{M}_k\) has the same dimensions as \(\text{slice}_k\) and assigns each pixel to its most likely anatomical class (or background). 

A key challenge in this setting is that only a small fraction of slices contain ground-truth annotations, making it impractical to fully train or fine-tune a conventional supervised model. Instead, we leverage a large-scale pretrained segmentation model \(f\). By providing a small number of labeled slices in \(S\), \(f\) can perform few-shot segmentation with high accuracy across unlabeled slices.

\subsection{UniverSeg}
UniverSeg \cite{butoi_universeg_2023} is a versatile model designed for few-shot segmentation in medical imaging. It is trained on 53 datasets, totaling over 22,000 CT and MR examinations. Crucially, UniverSeg does not require additional training when adapting to a new segmentation task; rather, it conditions on a small set of labeled images, referred to as the support set.

\begin{figure}[t]
    \centering
    \includegraphics[width=0.8\textwidth]{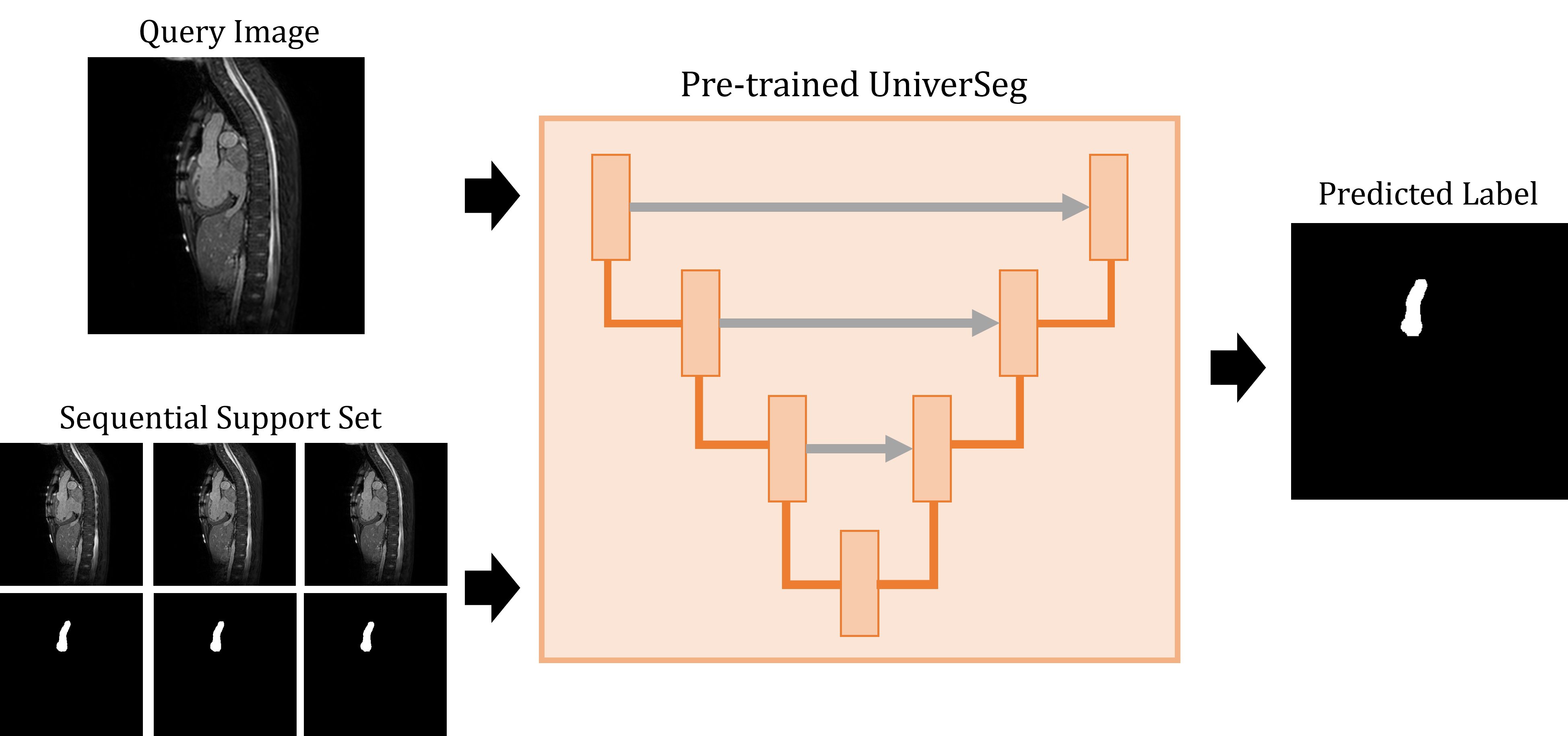} 
    \caption{UniverSeg for sequential inference} 
    \label{universeg} 
\end{figure}

Architecturally, UniverSeg adopts an encoder–decoder structure and uses “CrossBlock” modules for bidirectional feature exchange between query images and support images. Similar to U-Net \cite{ronneberger_u-net_2015}, it extracts features at multiple scales via the encoder and then reconstructs segmentation masks in the decoder. Pretrained on the large-scale MegaMedical dataset, UniverSeg has shown strong generalization across different tasks and modalities. In this study, we supply consecutive labeled slices as the support set and infer masks for the remaining slices (Figure \ref{universeg}). Although the standard UniverSeg pipeline processes each slice independently, it does not guarantee spatial consistency across neighboring slices. To address this limitation, we introduce an extension in the following section.

\begin{figure}[t]
    \centering
    \includegraphics[width=0.8\textwidth]{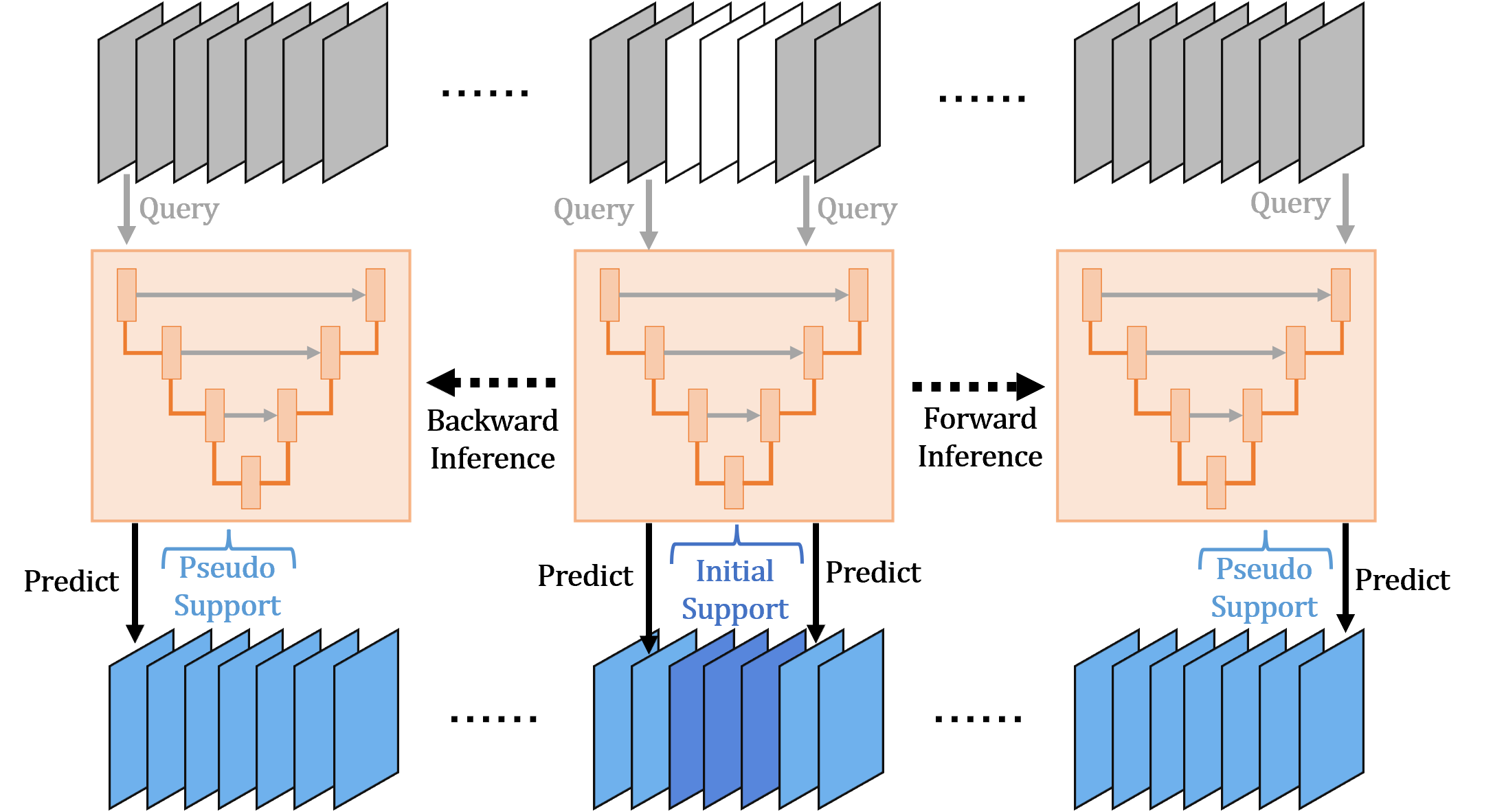} 
    \caption{An overview of in-contest cascade segmentation} 
    \label{ics} 
\end{figure}

\subsection{In-context Cascade Segmentation (ICS)}
We propose in-context cascade segmentation (ICS), an approach built on UniverSeg for sequential segmentation of consecutive slices (Figure \ref{ics}). ICS enhances consistency by incorporating each newly inferred slice into the support set for subsequent inferences, thus propagating information both forward and backward through the image volume.

While Algorithm 1 details the procedure, the ICS workflow can be summarized as follows:
\begin{enumerate}
    \item Initialization of the Support Set
    
    Select a small number of labeled slices near the center of the volume or from anatomically critical regions to form the initial support set.

    \item Bidirectional Sequential Inference
  
    From this initial set, apply UniverSeg to perform inference in both forward (increasing slice index) and backward (decreasing slice index) directions. After each slice is inferred, add the predicted mask to the support set for the next slice. To prevent the support set from growing unbounded, retain only the most recent \(m\) slices, discarding older entries.
    
\end{enumerate}
During each inference, data augmentation is applied to the support images to improve robustness. By iteratively updating the support set, ICS maintains a running repository of inferred masks and labeled slices. This process stabilizes segmentation quality across slices, especially for boundaries and anatomically intricate regions. Because the model infers each slice in sequence, it effectively uses prior inferences as a form of context, improving slice-to-slice consistency relative to straightforward, independent inferences by UniverSeg. 

\begin{algorithm}[t]
\caption{In-context Cascade Segmentation (ICS)}
\begin{algorithmic}[1]
\Require Volume $V = \{slice_1, slice_2, \dots, slice_n\}$
\Require InitialSupportSet = $\{(img_{c_1}, label_{c_1}), \dots, (img_{c_k}, label_{c_k})\}$
\Require Model UniverSeg
\Require Maximum support set size $m$
\Require Augmentation function $Augment()$
\Ensure Segmented volume with label predictions for each slice in $V$

\State Initialize $ForwardSupportSet \gets InitialSupportSet$
\State Initialize $BackwardSupportSet \gets InitialSupportSet$
\State Initialize $ForwardPredictedLabels \gets []$
\State Initialize $BackwardPredictedLabels \gets []$

\For{$i = \max(c_1, c_2, \dots, c_k)$ to $n$} 
    \State $QueryImage \gets slice_i$
    \State $PredictedLabel \gets UniverSeg(QueryImage, Augment(ForwardSupportSet))$
    \State $ForwardPredictedLabels \gets \text{Append}(ForwardPredictedLabels, PredictedLabel)$
    \State $ForwardSupportSet \gets \text{Append}(ForwardSupportSet, (slice_i, PredictedLabel))$
    \If{$|ForwardSupportSet| > m$}
        \State Remove the oldest entries so that $|ForwardSupportSet| = m$
    \EndIf
\EndFor

\For{$i = \min(c_1, c_2, \dots, c_k)$ down to $1$} 
    \State $QueryImage \gets slice_i$
    \State $PredictedLabel \gets UniverSeg(QueryImage, Augment(BackwardSupportSet))$
    \State $BackwardPredictedLabels \gets \text{Append}(BackwardPredictedLabels, PredictedLabel)$
    \State $BackwardSupportSet \gets \text{Append}(BackwardSupportSet, (slice_i, PredictedLabel))$
    \If{$|BackwardSupportSet| > m$}
        \State Remove the oldest entries so that $|BackwardSupportSet| = m$
    \EndIf
\EndFor

\State \Return $ForwardPredictedLabels$, $BackwardPredictedLabels$
\end{algorithmic}
\end{algorithm}

\section{Experiments}
\subsection{Dataset}
We evaluate our proposed approach using the HVSMR dataset \cite{pace_hvsmr-20_2024}, which contains 60 cardiovascular MRI scans from Boston Children’s Hospital. These scans include segmentation masks for four cardiac chambers (left ventricle (LV), right ventricle (RV), left atrium (LA), and right atrium (RA)) and four major vessels (ascending aorta (AO), pulmonary artery (PA), superior vena cava (SVC), and inferior vena cava (IVC)). Ground-truth segmentation masks were generated by combining manual annotations with automated outputs from a 3D-UNet, followed by expert revision. Each case encompasses a variety of congenital heart conditions, as well as surgical histories, resulting in diverse anatomical structures.

Since this study does not involve parameter tuning through training, we used all cases to evaluate the proposed method. In addition, each of the eight anatomical regions was treated as a separate dataset. As a preprocessing step, we removed any slices that did not contain the region of interest (i.e., no annotated labels).

\subsection{Experimental Settings}
We conduct three sets of experiments to investigate the performance and characteristics of the proposed method:

\subsubsection*{Quantitative and Qualitative Comparison to Baseline}
We compare ICS against the baseline UniverSeg approach (without in-context updates) using identical initial support slices and augmentation settings. We measure overall accuracy and inspect segmentation quality, particularly for difficult anatomical boundaries.

\subsubsection*{Effect of Varying the Number of Support Slices}
We assess how the number of initially labeled slices 
(\(m\)) influences segmentation performance by varying 
\(m\) from 1 to 5. This experiment clarifies whether adding more support slices consistently improves accuracy or if diminishing returns emerge.

\subsubsection*{Effect of Initial Support Slice Positions}
We investigate how the spatial location of the initial labeled slices within the volume affects segmentation outcomes. Specifically, for each anatomical target, we vary the position of the initial labeled slices while keeping the number of initial slices fixed.

For visualization or in-depth discussion, some results focus on $\mathit Patient\_32$, randomly chosen from the dataset.

\subsection{Evaluation Metrics}
We use the Dice similarity coefficient (DSC), as our primary metric for quantitative evaluation. The DSC is defined as:
\[
\text{DSC} = \frac{2 \cdot \text{TP}}{2 \cdot \text{TP} + \text{FP} + \text{FN}}
\]
where TP (True Positive), FP (False Positive), and FN (False Negative) are computed on a pixel-by-pixel basis, comparing predicted segmentation masks to ground truth. This metric is widely used for measuring overlap between two label sets.

\subsection{Implementation}
All experiments were conducted using Python 3.10.8. We employed the publicly available UniverSeg pretrained model \cite{butoi_universeg_2023} on a system equipped with an Intel(R) Core(TM) i9-10940X CPU (3.30 GHz) and a Quadro RTX 8000 GPU (48 GB). In our ICS procedure, data augmentation is limited to rotations of the support slices by 90°, 180°, and 270°. Unless otherwise stated, the support set size \(m\) is fixed at 5, and any older slices are discarded once the support set grows beyond \(m\).

\section{Results}
\begin{figure}[t]
    \centering
    \includegraphics[width=0.8\textwidth]{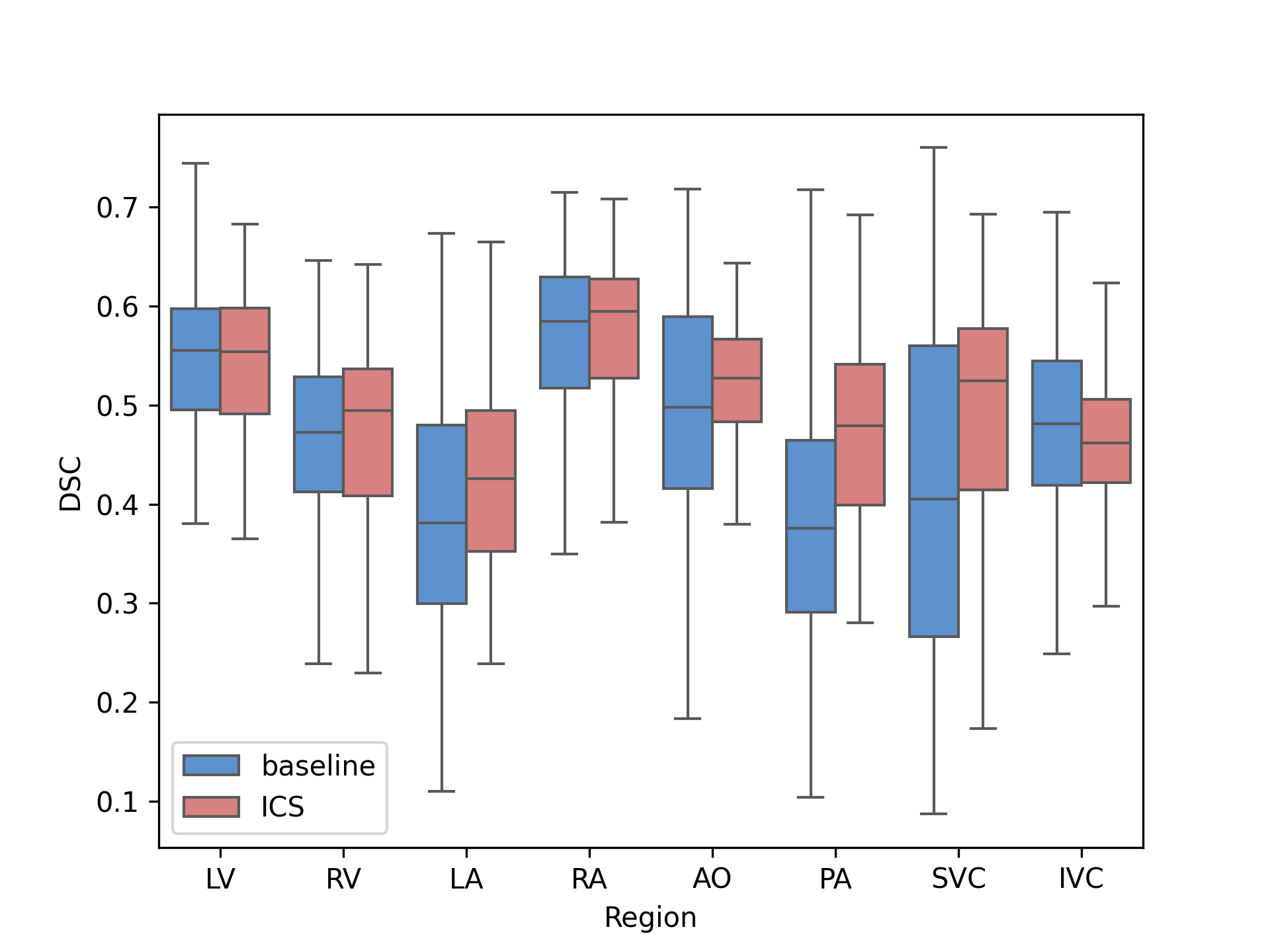} 
    \caption{Box plots of the DSC for the baseline method (green) and ICS (orange) across each anatomical region. The boxes represent the interquartile range and the horizontal line within each box indicates the median.} 
    \label{m5_boxplot} 
\end{figure}

First, the comparison with the baseline method appeared in Table \ref{comparison} and Figure \ref{m5_boxplot}.
As shown in \ref{comparison}, ICS achieved significantly higher DSC (p<0.05) than the baseline method in the LA, RA, AO, PA, and SVC regions, whereas no significant differences were observed in the LV, RV, and IVC regions.

\begin{table}[b]
\centering
\caption{Comparison of DSC statistics (mean and standard deviation) between the baseline method and ICS for each anatomical region in the HVSMR dataset. Bold values indicate regions where ICS achieved significantly higher performance than the baseline method (p < 0.05).}
\label{comparison}
\begin{tabular}{lccccc}
\toprule
\multirow{2}{*}{Region} & \multicolumn{2}{c}{Baseline} & \multicolumn{2}{c}{ICS} & \multirow{2}{*}{p-value} \\
\cmidrule(lr){2-3}\cmidrule(lr){4-5}
 & Mean & Std & Mean & Std & \\
\midrule
LV  & 0.5489 & 0.0893 & 0.5475 & 0.0770 & 0.8903 \\  
RV  & 0.4663 & 0.1047 & 0.4741 & 0.0933 & 0.3820 \\  
LA  & 0.3879 & 0.1210 & \textbf{0.4272} & 0.0925 & 0.0007 \\  
RA  & 0.5500 & 0.1191 & \textbf{0.5734} & 0.0791 & 0.0214 \\  
AO  & 0.4945 & 0.1269 & \textbf{0.5210} & 0.0679 & 0.0206 \\  
PA  & 0.3807 & 0.1336 & \textbf{0.4745} & 0.0974 & <0.0001 \\  
SVC & 0.4079 & 0.1805 & \textbf{0.4938} & 0.1226 & <0.0001 \\  
IVC & 0.4607 & 0.1258 & 0.4629 & 0.0871 & 0.8561 \\  
\bottomrule
\end{tabular}
\end{table}

\begin{figure}[t]
    \begin{minipage}{\textwidth}
        \hspace{-2.3cm} 
        \includegraphics[width=1.26\textwidth]{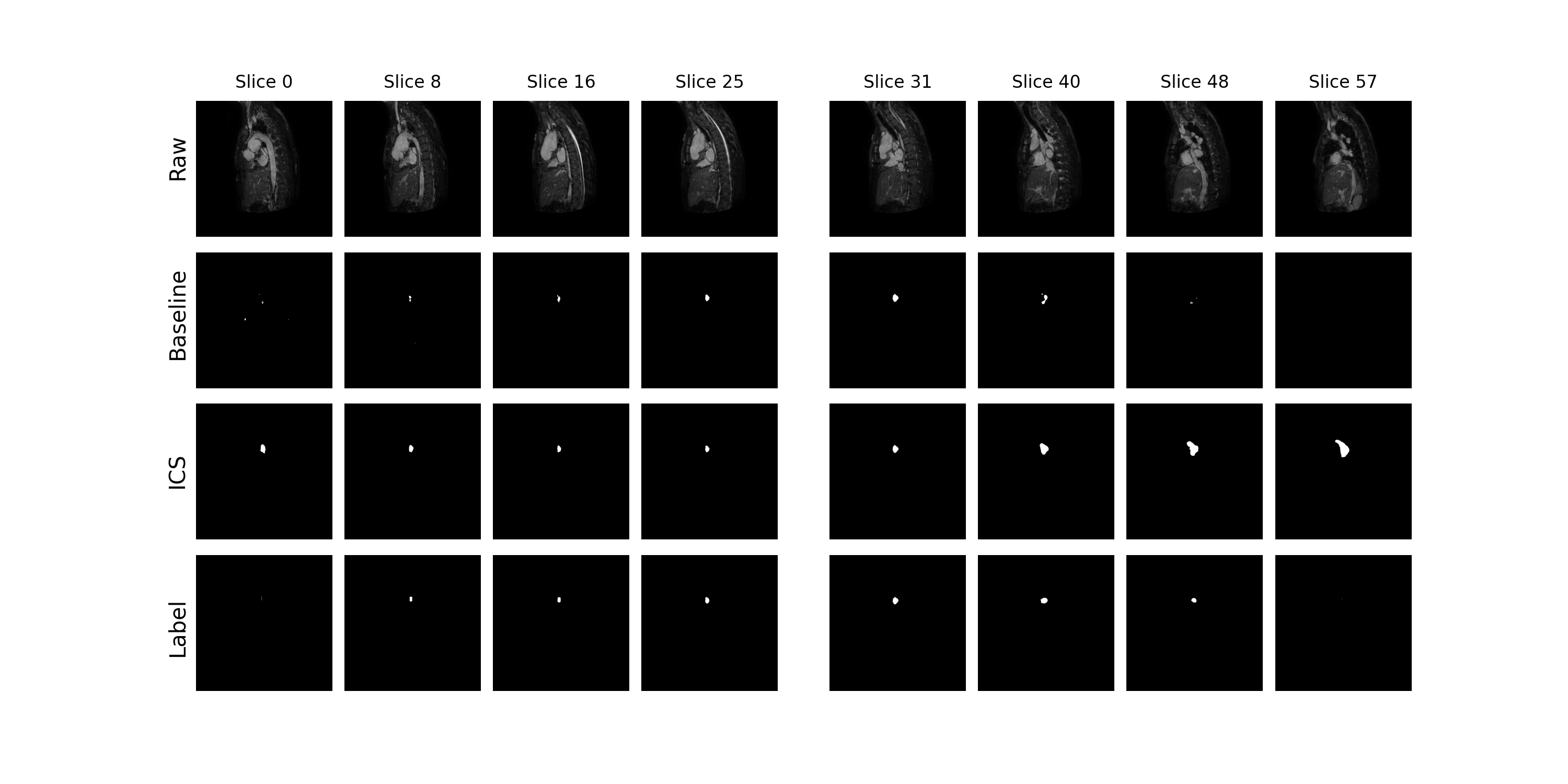} 
    \end{minipage}
    
    \caption{Segmentation results for the PA region in selected slices of a patient’s volume. From top to bottom: the raw image, the prediction by the baseline method, the prediction by ICS, and the ground truth label. Columns show different slice positions.} 
    \label{PA_quality} 
\end{figure}

\begin{figure}[t]
    \begin{minipage}{\textwidth}
        \hspace{-2.3cm}
        \includegraphics[width=1.26\textwidth]{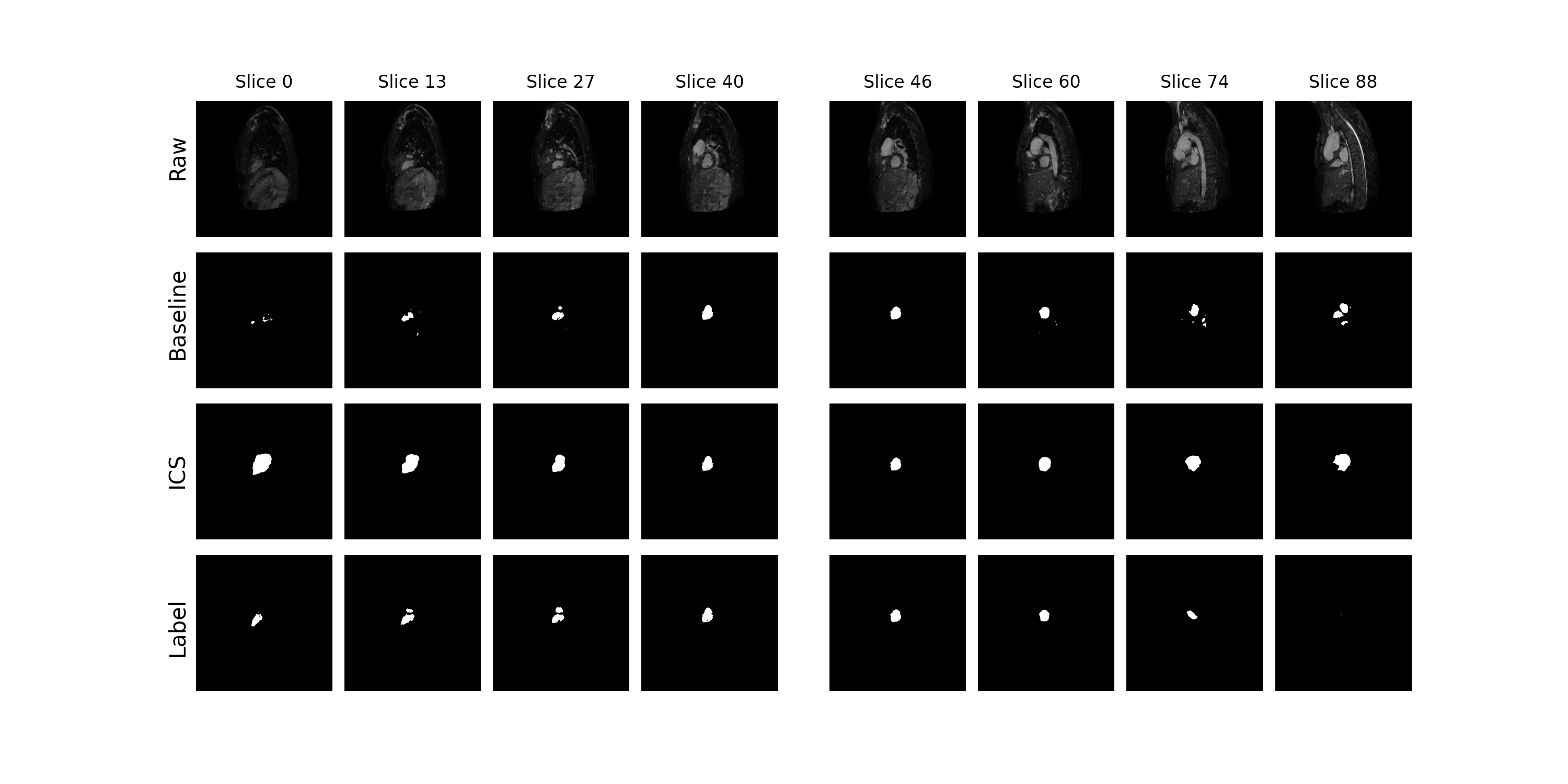} 
    \end{minipage}
    
    \caption{Segmentation results for the LV region in selected slices of a patient’s volume. From top to bottom: the raw image, the prediction by the baseline method, the prediction by ICS, and the ground truth label. Columns show different slice positions.} 
    \label{LV_quality} 
\end{figure}

Figure \ref{PA_quality} illustrates a case in which ICS significantly outperformed the baseline method for the PA region of $Patient\_32$. With the baseline method, fine structures and positional changes were not adequately captured, leading to discontinuous predictions and extensive missing areas. By contrast, ICS consistently identified the target region across all slices and produced stable masks.
Figure \ref{LV_quality} shows a case where the baseline method surpassed ICS for the LV region of the same patient. As the slices progressed to the right, anatomical features shifted; although the baseline method sometimes introduced noise, certain slices were delineated more accurately than by ICS. Meanwhile, although ICS generally produced stable masks, it tended to over-segment the target region, resulting in more false positives.

\begin{figure}[t]
    \hspace{-1cm}
    \includegraphics[width=1.1\textwidth]{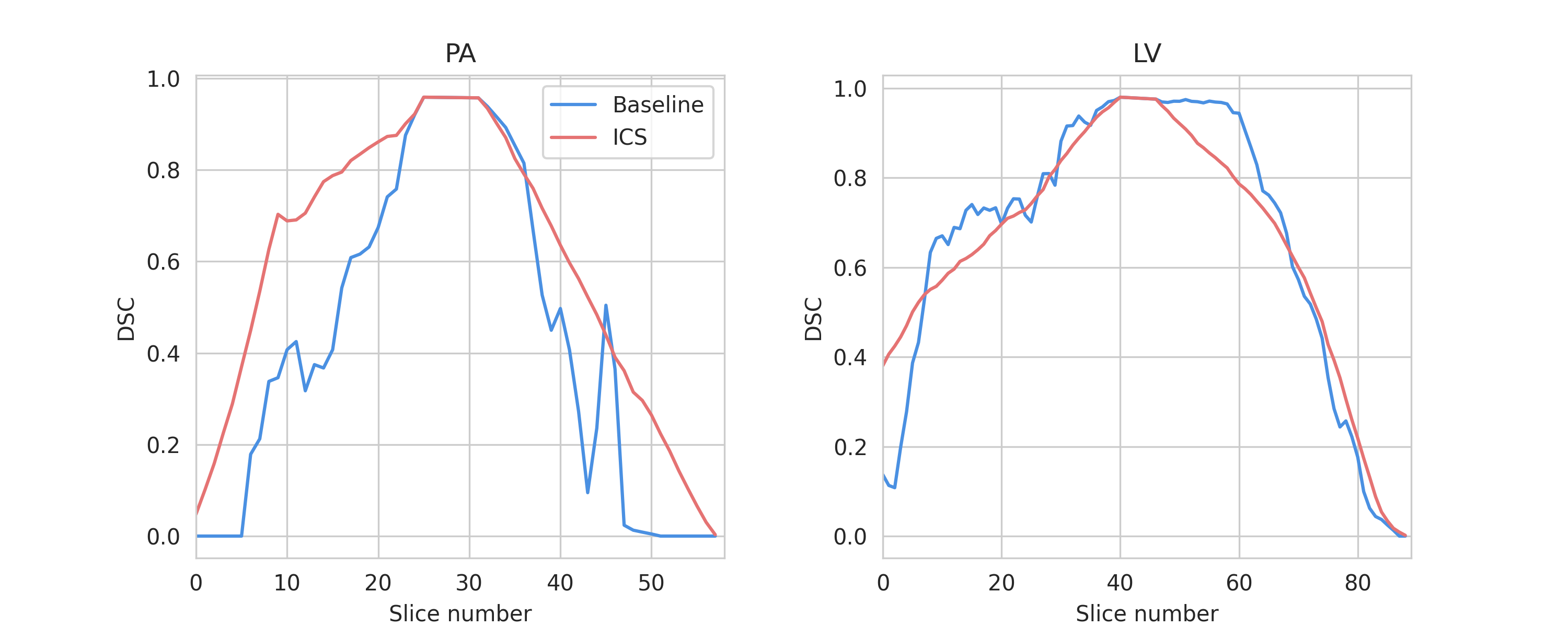} 
    \caption{Line plots of the DSC for the baseline method (blue) and ICS (red) across consecutive slices in the PA (left) and LV (right) regions of a selected patient. The horizontal axis indicates the slice number.} 
    \label{dice_per_slice} 
\end{figure}

Figure \ref{dice_per_slice} shows the slice-by-slice DSC for LV and PA in $Patient\_32$, offering a quantitative perspective of the patterns noted in Figures \ref{PA_quality} and \ref{LV_quality}. In both regions, ICS exhibits less variation between slices compared to the baseline method.

\begin{figure}[t]
    \centering
    \includegraphics[width=0.9\textwidth]{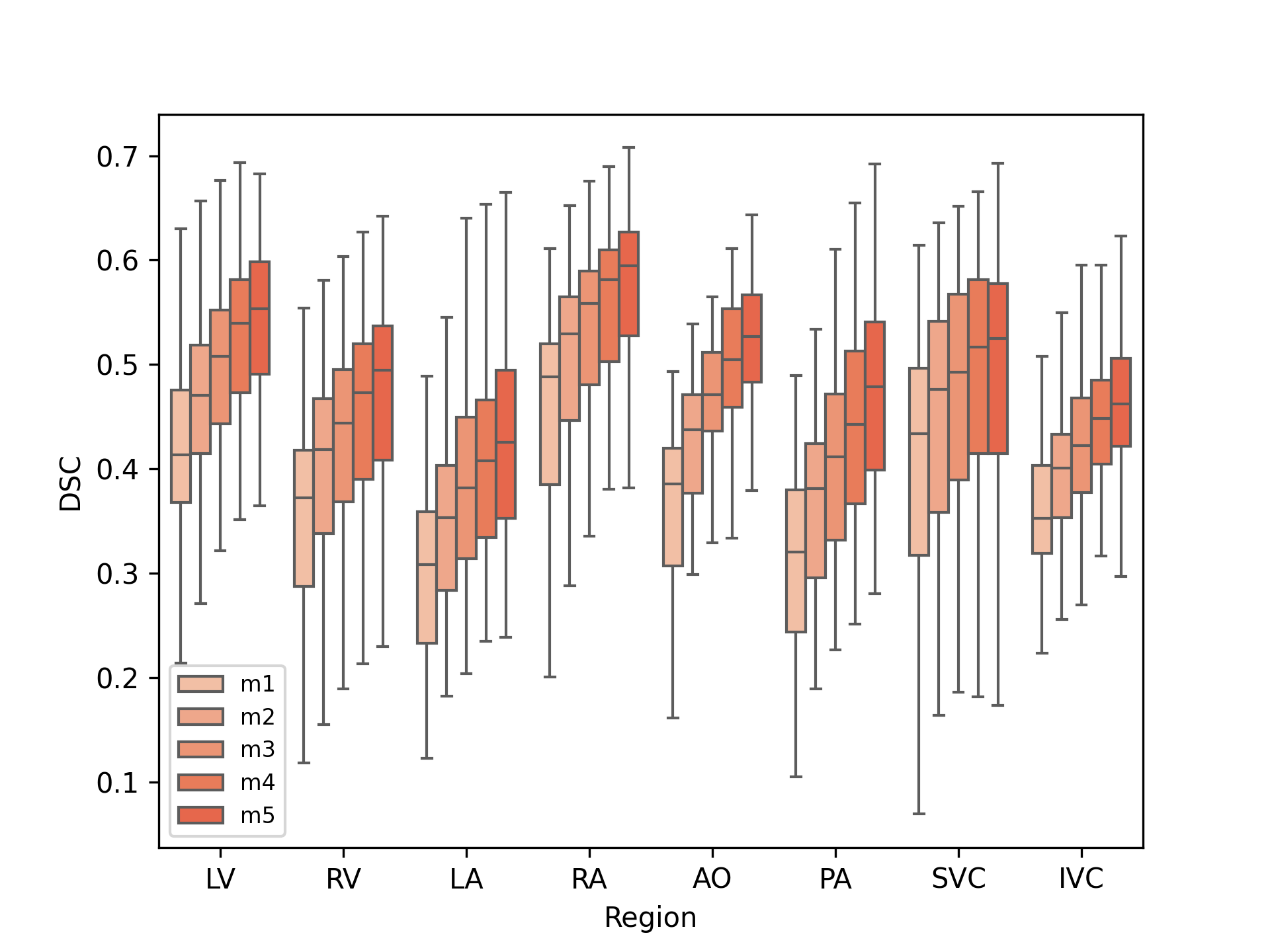} 
    \caption{Box plots of the DSC for each anatomical region when varying the number of initial labeled slices \( m \). Each color corresponds to a different \( m \) setting (from \( m = 1\) to \( m = 5\)).} 
    \label{m1_m5} 
\end{figure}

Figure \ref{m1_m5} depicts the results of an experiment varying the number of initial support slices 
 \( m \). As  \( m \) increases, all anatomical regions display an upward trend in DSC.

\begin{figure}[t]
    \centering
    \includegraphics[width=1.0\textwidth]{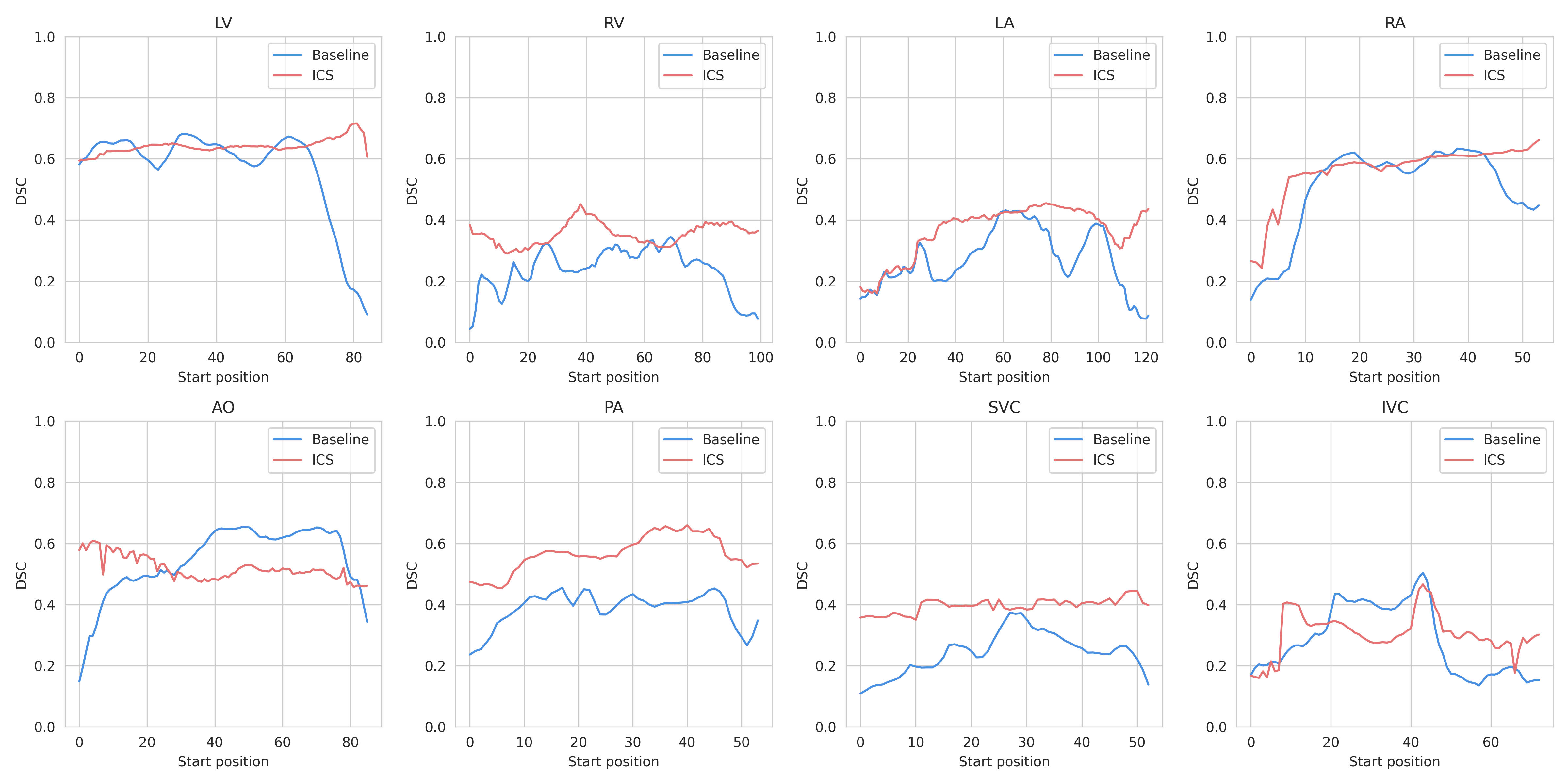} 
    \caption{Line plots of the DSC for each anatomical region when varying the initial labeled slice position.} 
    \label{location_change} 
\end{figure}

Finally, Figure \ref{location_change} demonstrates the the effect of altering the initial labeled slice positions for each region. The results indicate that these positions strongly influence segmentation accuracy. Moreover, while ICS consistently outperforms the baseline method for PA and SVC regardless of initial placement, it may perform worse than the baseline in other regions, depending on how the initial slices are selected.

\section{Discussion}
This study presented three experiments to assess the effectiveness and characteristics of ICS. Based on these findings, we offer the following observations, followed by a discussion of the study’s limitations.

ICS demonstrated significantly higher DSC values compared to the baseline method (UniverSeg) in the regions of LA, RA, AO, PA, and SVC (Table \ref{comparison}). 
As exemplified in Figure \ref{PA_quality}, ICS showed especially strong performance when maintaining slice-to-slice consistency and capturing anatomically complex structures, such as the PA. On the other hand, ICS did not outperform the baseline in the LV and IVC regions, displaying a tendency toward over-segmentation in some cases (Figure \ref{LV_quality}). This tendency, which leads to increased false positives, resembles behaviors reported in self-training–based methods \cite{takaya_sequential_2021}. Introducing a mechanism to incorporate confidence metrics or additional constraints during support-set updates could mitigate this over-segmentation.

In the experiment varying the number of initial support slices \( m \), most regions showed an improvement in DSC as \( m \) increased (Figure \ref{m1_m5}). Particularly when a sufficient number of support slices were provided, ICS achieved stable segmentation results. This trend is consistent with observations from the original UniverSeg \cite{butoi_universeg_2023}. However, from a computational standpoint, relying on a larger number of support slices raises concerns about increased inference time and GPU memory usage. These findings suggest that selecting an appropriate number of initial support slices is crucial in situations with limited computational resources. Alternatively, model distillation \cite{gou_knowledge_2021} could be employed to optimize the efficiency of the UniverSeg model itself, alongside other strategies for improving computational efficiency.

The experiment varying the position of the initial support slices revealed that the position has a significant impact on segmentation accuracy. The DSC plot for each slice in Figure \ref{dice_per_slice} shows a trend where the deviation from the ground truth increases as the distance from the initial position grows. While this suggests that the initial position should ideally be set at the center of the region of interest, it is noteworthy that this is not necessarily the case for all anatomical regions, which is an intriguing finding.

Future developments should focus on establishing methods to determine appropriate initial slices before annotation. This challenge aligns with the cold-start problem in active learning \cite{jin_cold-start_2022}\cite{kazeminia_data-driven_2024}, and various approaches, such as clustering-based methods, could potentially be leveraged to address it.

The limitations of this study are outlined in two points.
The first limitation is the assumption that all volumes, including the boundary slices, are labeled. In real-world data, slices that do not contain the region of interest may be present, necessitating a mechanism to appropriately stop label propagation for such slices.
The second limitation is the lack of validation datasets. Because UniverSeg utilizes a large portion of publicly available datasets, it was necessary to rely on the limited open datasets that were not used for pre-training. Consequently, this study was effectively restricted to using only the HVSMR dataset. Future work should include in-house datasets and extend evaluation to various modalities, such as CT and ultrasound, to further validate the method.

\section{Conclusion}
This study proposed the In-context Cascade Segmentation (ICS) method to improve segmentation performance in sequential medical images with minimal annotations. By leveraging the UniverSeg framework, ICS sequentially updates its support set using inference results, propagating segmentation information across slices in both forward and backward directions. Through experiments on the HVSMR dataset, ICS demonstrated significant improvements over the baseline method in several anatomical regions, particularly in ensuring inter-slice consistency and accurately capturing complex structures.

The results highlighted that ICS performs well when a sufficient number of initial support slices are provided and that the position of these slices significantly impacts the segmentation performance. However, challenges such as over-segmentation in certain regions, the computational cost associated with a large number of support slices, and the need for effective initial slice selection were also identified. These findings suggest several avenues for future work, including the optimization of support set updates, model efficiency improvements, and the development of automated strategies for selecting optimal initial slices.

In conclusion, ICS represents a promising step toward reducing annotation burdens in medical image segmentation while maintaining high performance and consistency. Further validation using diverse datasets and imaging modalities will help establish its robustness and broaden its applicability in clinical and research settings.

\section*{Acknowledgments}
This work was supported by JSPS KAKENHI Grant Number JP24K15174.
\bibliographystyle{unsrt}  
\bibliography{ics}

\begin{thebibliography}{10}

\bibitem{mall_comprehensive_2023}
Pawan~Kumar Mall, Pradeep~Kumar Singh, Swapnita Srivastav, Vipul Narayan, Marcin Paprzycki, Tatiana Jaworska, and Maria Ganzha.
\newblock A comprehensive review of deep neural networks for medical image processing: {Recent} developments and future opportunities.
\newblock {\em Healthcare Analytics}, 4:100216, December 2023.

\bibitem{azad_medical_2024}
Reza Azad, Ehsan~Khodapanah Aghdam, Amelie Rauland, Yiwei Jia, Atlas~Haddadi Avval, Afshin Bozorgpour, Sanaz Karimijafarbigloo, Joseph~Paul Cohen, Ehsan Adeli, and Dorit Merhof.
\newblock Medical {Image} {Segmentation} {Review}: {The} {Success} of {U}-{Net}.
\newblock {\em IEEE Transactions on Pattern Analysis and Machine Intelligence}, 46(12):10076--10095, December 2024.

\bibitem{dot_fully_2022}
Gauthier Dot, Thomas Schouman, Guillaume Dubois, Philippe Rouch, and Laurent Gajny.
\newblock Fully automatic segmentation of craniomaxillofacial {CT} scans for computer-assisted orthognathic surgery planning using the {nnU}-{Net} framework.
\newblock {\em European Radiology}, 32(6):3639--3648, June 2022.

\bibitem{harrison_machine_2022}
K.~Harrison, H.~Pullen, C.~Welsh, O.~Oktay, J.~Alvarez-Valle, and R.~Jena.
\newblock Machine {Learning} for {Auto}-{Segmentation} in {Radiotherapy} {Planning}.
\newblock {\em Clinical Oncology}, 34(2):74--88, February 2022.

\bibitem{zhang_deeprecs_2022}
Yue Zhang, Chengtao Peng, Liying Peng, Yingying Xu, Lanfen Lin, Ruofeng Tong, Zhiyi Peng, Xiongwei Mao, Hongjie Hu, Yen-Wei Chen, and Jingsong Li.
\newblock {DeepRecS}: {From} {RECIST} {Diameters} to {Precise} {Liver} {Tumor} {Segmentation}.
\newblock {\em IEEE Journal of Biomedical and Health Informatics}, 26(2):614--625, February 2022.

\bibitem{shimokawa_deep_2023}
Daiki Shimokawa, Kengo Takahashi, Ken Oba, Eichi Takaya, Takuma Usuzaki, Mizuki Kadowaki, Kurara Kawaguchi, Maki Adachi, Tomofumi Kaneno, Toshinori Fukuda, Kazuyo Yagishita, Hiroko Tsunoda, and Takuya Ueda.
\newblock Deep learning model for predicting the presence of stromal invasion of breast cancer on digital breast tomosynthesis.
\newblock {\em Radiological Physics and Technology}, 16(3):406--413, September 2023.

\bibitem{haraguchi_radiomics_2023}
Takafumi Haraguchi, Yasuyuki Kobayashi, Daisuke Hirahara, Tatsuaki Kobayashi, Eichi Takaya, Mariko~Takishita Nagai, Hayato Tomita, Jun Okamoto, Yoshihide Kanemaki, and Koichiro Tsugawa.
\newblock Radiomics model of diffusion-weighted whole-body imaging with background signal suppression ({DWIBS}) for predicting axillary lymph node status in breast cancer.
\newblock {\em Journal of X-Ray Science and Technology}, 31(3):627--640, 2023.

\bibitem{ronneberger_u-net_2015}
Olaf Ronneberger, Philipp Fischer, and Thomas Brox.
\newblock U-{Net}: {Convolutional} {Networks} for {Biomedical} {Image} {Segmentation}.
\newblock In {\em Medical image computing and computer-assisted intervention, {Lecture} {Notes} in {Computer} {Science}}, volume 9351. Springer Cham, 2015.

\bibitem{cao_swin-unet_2022}
Hu~Cao, Yueyue Wang, Joy Chen, Dongsheng Jiang, Xiaopeng Zhang, Qi~Tian, and Manning Wang.
\newblock Swin-unet: {Unet}-like pure transformer for medical image segmentation.
\newblock In {\em European conference on computer vision}, pages 205--218. Springer, 2022.

\bibitem{das_attention-unet_2024}
Niharika Das and Sujoy Das.
\newblock Attention-{UNet} architectures with pretrained backbones for multi-class cardiac {MR} image segmentation.
\newblock {\em Current Problems in Cardiology}, 49(1):102129, January 2024.

\bibitem{mei_radimagenet_2022}
Xueyan Mei, Zelong Liu, Philip~M. Robson, Brett Marinelli, Mingqian Huang, Amish Doshi, Adam Jacobi, Chendi Cao, Katherine~E. Link, Thomas Yang, Ying Wang, Hayit Greenspan, Timothy Deyer, Zahi~A. Fayad, and Yang Yang.
\newblock {RadImageNet}: {An} {Open} {Radiologic} {Deep} {Learning} {Research} {Dataset} for {Effective} {Transfer} {Learning}.
\newblock {\em Radiology: Artificial Intelligence}, 4(5):e210315, September 2022.

\bibitem{xu_vis-mae_2025}
Zelong Liu, Andrew Tieu, Nikhil Patel, George Soultanidis, Louisa Deyer, Ying Wang, Sean Huver, Alexander Zhou, Yunhao Mei, Zahi~A. Fayad, Timothy Deyer, and Xueyan Mei.
\newblock {VIS}-{MAE}: {An} {Efficient} {Self}-supervised {Learning} {Approach} on {Medical} {Image} {Segmentation} and {Classification}.
\newblock In Xuanang Xu, Zhiming Cui, Islem Rekik, Xi~Ouyang, and Kaicong Sun, editors, {\em Machine {Learning} in {Medical} {Imaging}}, volume 15242, pages 95--107. Springer Nature Switzerland, Cham, 2025.
\newblock Series Title: Lecture Notes in Computer Science.

\bibitem{wu_medical_2023}
Junde Wu, Wei Ji, Yuanpei Liu, Huazhu Fu, Min Xu, Yanwu Xu, and Yueming Jin.
\newblock Medical {SAM} {Adapter}: {Adapting} {Segment} {Anything} {Model} for {Medical} {Image} {Segmentation}, December 2023.
\newblock arXiv:2304.12620 [cs].

\bibitem{zhu_medical_2024}
Jiayuan Zhu, Abdullah Hamdi, Yunli Qi, Yueming Jin, and Junde Wu.
\newblock Medical {SAM} 2: {Segment} medical images as video via {Segment} {Anything} {Model} 2, December 2024.
\newblock arXiv:2408.00874 [cs].

\bibitem{ouyang_causality-inspired_2023}
Cheng Ouyang, Chen Chen, Surui Li, Zeju Li, Chen Qin, Wenjia Bai, and Daniel Rueckert.
\newblock Causality-{Inspired} {Single}-{Source} {Domain} {Generalization} for {Medical} {Image} {Segmentation}.
\newblock {\em IEEE Transactions on Medical Imaging}, 42(4):1095--1106, April 2023.

\bibitem{guan_domain_2022}
Hao Guan and Mingxia Liu.
\newblock Domain {Adaptation} for {Medical} {Image} {Analysis}: {A} {Survey}.
\newblock {\em IEEE Transactions on Biomedical Engineering}, 69(3):1173--1185, March 2022.

\bibitem{zhang_generalizing_2020}
Ling Zhang, Xiaosong Wang, Dong Yang, Thomas Sanford, Stephanie Harmon, Baris Turkbey, Bradford~J. Wood, Holger Roth, Andriy Myronenko, Daguang Xu, and Ziyue Xu.
\newblock Generalizing {Deep} {Learning} for {Medical} {Image} {Segmentation} to {Unseen} {Domains} via {Deep} {Stacked} {Transformation}.
\newblock {\em IEEE Transactions on Medical Imaging}, 39(7):2531--2540, July 2020.

\bibitem{hasani_artificial_2022}
Navid Hasani, Faraz Farhadi, Michael~A. Morris, Moozhan Nikpanah, Arman Rahmim, Yanji Xu, Anne Pariser, Michael~T. Collins, Ronald~M. Summers, Elizabeth Jones, Eliot Siegel, and Babak Saboury.
\newblock Artificial {Intelligence} in {Medical} {Imaging} and its {Impact} on the {Rare} {Disease} {Community}: {Threats}, {Challenges} and {Opportunities}.
\newblock {\em PET Clinics}, 17(1):13--29, January 2022.

\bibitem{butoi_universeg_2023}
Victor~Ion Butoi, Jose~Javier Gonzalez~Ortiz, Tianyu Ma, Mert~R. Sabuncu, John Guttag, and Adrian~V. Dalca.
\newblock {UniverSeg}: {Universal} {Medical} {Image} {Segmentation}.
\newblock In {\em 2023 {IEEE}/{CVF} {International} {Conference} on {Computer} {Vision} ({ICCV})}, pages 21381--21394, Paris, France, October 2023. IEEE.

\bibitem{pace_hvsmr-20_2024}
Danielle~F. Pace, Hannah T.~M. Contreras, Jennifer Romanowicz, Shruti Ghelani, Imon Rahaman, Yue Zhang, Patricia Gao, Mohammad~Imrul Jubair, Tom Yeh, Polina Golland, Tal Geva, Sunil Ghelani, Andrew~J. Powell, and Mehdi~Hedjazi Moghari.
\newblock {HVSMR}-2.0: {A} {3D} cardiovascular {MR} dataset for whole-heart segmentation in congenital heart disease.
\newblock {\em Scientific Data}, 11(1):721, July 2024.

\bibitem{pham_current_2000}
Dzung~L Pham, Chenyang Xu, and Jerry~L Prince.
\newblock Current methods in medical image segmentation.
\newblock {\em Annual review of biomedical engineering}, 2(1):315--337, 2000.
\newblock Publisher: Annual Reviews 4139 El Camino Way, PO Box 10139, Palo Alto, CA 94303-0139, USA.

\bibitem{ng_medical_2006}
HP~Ng, SH~Ong, KWC Foong, Poh-Sun Goh, and WL~Nowinski.
\newblock Medical image segmentation using k-means clustering and improved watershed algorithm.
\newblock In {\em 2006 {IEEE} southwest symposium on image analysis and interpretation}, pages 61--65. IEEE, 2006.

\bibitem{ng_masseter_2008}
H.P. Ng, S.H. Ong, K.W.C. Foong, P.S. Goh, and W.L. Nowinski.
\newblock Masseter segmentation using an improved watershed algorithm with unsupervised classification.
\newblock {\em Computers in Biology and Medicine}, 38(2):171--184, February 2008.

\bibitem{ricci_retinal_2007}
Elisa Ricci and Renzo Perfetti.
\newblock Retinal {Blood} {Vessel} {Segmentation} {Using} {Line} {Operators} and {Support} {Vector} {Classification}.
\newblock {\em IEEE Transactions on Medical Imaging}, 26(10):1357--1365, October 2007.

\bibitem{mahapatra_analyzing_2014}
Dwarikanath Mahapatra.
\newblock Analyzing {Training} {Information} {From} {Random} {Forests} for {Improved} {Image} {Segmentation}.
\newblock {\em IEEE Transactions on Image Processing}, 23(4):1504--1512, April 2014.

\bibitem{ciresan_deep_2012}
Dan~C Ciresan, Luca~M Gambardella, and Alessandro Giusti.
\newblock Deep {Neural} {Networks} {Segment} {Neuronal} {Membranes} in {Electron} {Microscopy} {Images}.
\newblock In {\em Proceedings of the 25th {Iternational} {Conference} on {Neural} {Information} {Processing} {Systems}}, pages 2843--2851, 2012.

\bibitem{chen_transunet_2024}
Jieneng Chen, Jieru Mei, Xianhang Li, Yongyi Lu, Qihang Yu, Qingyue Wei, Xiangde Luo, Yutong Xie, Ehsan Adeli, Yan Wang, Matthew~P. Lungren, Shaoting Zhang, Lei Xing, Le~Lu, Alan Yuille, and Yuyin Zhou.
\newblock {TransUNet}: {Rethinking} the {U}-{Net} architecture design for medical image segmentation through the lens of transformers.
\newblock {\em Medical Image Analysis}, 97:103280, October 2024.

\bibitem{milletari_v-net_2016}
Fausto Milletari, Nassir Navab, and Seyed-Ahmad Ahmadi.
\newblock V-net: {Fully} convolutional neural networks for volumetric medical image segmentation.
\newblock In {\em 2016 fourth international conference on {3D} vision ({3DV})}, pages 565--571. Ieee, 2016.

\bibitem{kirillov_segment_2023}
Alexander Kirillov, Eric Mintun, Nikhila Ravi, Hanzi Mao, Chloe Rolland, Laura Gustafson, Tete Xiao, Spencer Whitehead, Alexander~C. Berg, Wan-Yen Lo, Piotr Dollár, and Ross Girshick.
\newblock Segment {Anything}.
\newblock In {\em 2023 {IEEE}/{CVF} {International} {Conference} on {Computer} {Vision} ({ICCV})}, pages 3992--4003, Paris, France, October 2023. IEEE.

\bibitem{dong_survey_2024}
Qingxiu Dong, Lei Li, Damai Dai, Ce~Zheng, Jingyuan Ma, Rui Li, Heming Xia, Jingjing Xu, Zhiyong Wu, Tianyu Liu, Baobao Chang, Xu~Sun, Lei Li, and Zhifang Sui.
\newblock A {Survey} on {In}-context {Learning}, October 2024.
\newblock arXiv:2301.00234 [cs].

\bibitem{min_rethinking_2022}
Sewon Min, Xinxi Lyu, Ari Holtzman, Mikel Artetxe, Mike Lewis, Hannaneh Hajishirzi, and Luke Zettlemoyer.
\newblock Rethinking the {Role} of {Demonstrations}: {What} {Makes} {In}-{Context} {Learning} {Work}?, October 2022.
\newblock arXiv:2202.12837 [cs].

\bibitem{brown_language_2020}
Tom Brown, Benjamin Mann, Nick Ryder, Melanie Subbiah, Jared~D Kaplan, Prafulla Dhariwal, Arvind Neelakantan, Pranav Shyam, Girish Sastry, Amanda Askell, and {others}.
\newblock Language models are few-shot learners.
\newblock {\em Advances in neural information processing systems}, 33:1877--1901, 2020.

\bibitem{oswald_transformers_2023}
Johannes~von Oswald, Eyvind Niklasson, Ettore Randazzo, João Sacramento, Alexander Mordvintsev, Andrey Zhmoginov, and Max Vladymyrov.
\newblock Transformers learn in-context by gradient descent, May 2023.
\newblock arXiv:2212.07677 [cs].

\bibitem{zhang_what_2023}
Yuanhan Zhang, Kaiyang Zhou, and Ziwei Liu.
\newblock What makes good examples for visual in-context learning?
\newblock {\em Advances in Neural Information Processing Systems}, 36:17773--17794, 2023.

\bibitem{van_engelen_survey_2020}
Jesper~E. van Engelen and Holger~H. Hoos.
\newblock A survey on semi-supervised learning.
\newblock {\em Machine Learning}, 109(2):373--440, February 2020.

\bibitem{jiao_learning_2024}
Rushi Jiao, Yichi Zhang, Le~Ding, Bingsen Xue, Jicong Zhang, Rong Cai, and Cheng Jin.
\newblock Learning with limited annotations: {A} survey on deep semi-supervised learning for medical image segmentation.
\newblock {\em Computers in Biology and Medicine}, 169:107840, February 2024.

\bibitem{chaitanya_local_2023}
Krishna Chaitanya, Ertunc Erdil, Neerav Karani, and Ender Konukoglu.
\newblock Local contrastive loss with pseudo-label based self-training for semi-supervised medical image segmentation.
\newblock {\em Medical Image Analysis}, 87:102792, July 2023.

\bibitem{xia_uncertainty-aware_2020}
Yingda Xia, Dong Yang, Zhiding Yu, Fengze Liu, Jinzheng Cai, Lequan Yu, Zhuotun Zhu, Daguang Xu, Alan Yuille, and Holger Roth.
\newblock Uncertainty-aware multi-view co-training for semi-supervised medical image segmentation and domain adaptation.
\newblock {\em Medical Image Analysis}, 65:101766, October 2020.

\bibitem{wang_graphcl_2024}
Mengzhu Wang, Jiao Li, Houcheng Su, Nan Yin, Liang Yang, and Shen Li.
\newblock {GraphCL}: {Graph}-based {Clustering} for {Semi}-{Supervised} {Medical} {Image} {Segmentation}, November 2024.
\newblock arXiv:2411.13147 [cs].

\bibitem{takaya_sequential_2021}
Eichi Takaya, Yusuke Takeichi, Mamiko Ozaki, and Satoshi Kurihara.
\newblock Sequential semi-supervised segmentation for serial electron microscopy image with small number of labels.
\newblock {\em Journal of Neuroscience Methods}, 351:109066, March 2021.

\bibitem{gou_knowledge_2021}
Jianping Gou, Baosheng Yu, Stephen~J. Maybank, and Dacheng Tao.
\newblock Knowledge {Distillation}: {A} {Survey}.
\newblock {\em International Journal of Computer Vision}, 129(6):1789--1819, June 2021.

\bibitem{jin_cold-start_2022}
Qiuye Jin, Mingzhi Yuan, Shiman Li, Haoran Wang, Manning Wang, and Zhijian Song.
\newblock Cold-start active learning for image classification.
\newblock {\em Information Sciences}, 616:16--36, November 2022.

\bibitem{kazeminia_data-driven_2024}
Salome Kazeminia, Miroslav Březík, Sayedali~Shetab Boushehri, and Carsten Marr.
\newblock A {Data}-{Driven} {Solution} for {The} {Cold} {Start} {Problem} in {Biomedical} {Image} {Classification}.
\newblock In {\em 2024 {IEEE} {International} {Symposium} on {Biomedical} {Imaging} ({ISBI})}, pages 1--5, Athens, Greece, May 2024. IEEE.

\end{thebibliography}

\end{document}